\documentclass[trackchanges,twocolumn]{aastex701}

\usepackage{amsmath}
\usepackage{cancel}
\usepackage{placeins}
\usepackage{svg}
\usepackage{subscript}
\usepackage{makecell}
\usepackage{array} % REQUIRED for the m{...} and >{...} column types
\usepackage{subcaption}
\usepackage[multiple]{footmisc}
\shorttitle{Thermal radio emission and magnetic field strength in the `middle' corona}
\shortauthors{Sayuf, Kathiravan \& Ramesh}
\received{-}
\revised{-}
\accepted{-}
\begin{document}
	
	\title{\bf Estimates of magnetic field strength in the solar corona using observations of thermal radio emission at low frequencies}
	
	\author[orcid=0000-0003-4598-6830]{Shaik Sayuf}
	\affiliation{Indian Institute of Astrophysics, Koramangala 2nd Block, Bangalore 560034, Karnataka, India}
	\email[show]{shaik.sayufg5@gmail.com}
	\author[orcid=0000-0002-6126-8962]{C. Kathiravan}
	\affiliation{Indian Institute of Astrophysics, Koramangala 2nd Block, Bangalore 560034, Karnataka, India}
	\email{kathir@iiap.res.in}
	\author[orcid=0000-0003-2651-0204]{R. Ramesh}
	\affiliation{Indian Institute of Astrophysics, Koramangala 2nd Block, Bangalore 560034, Karnataka, India}
	\email{ramesh@iiap.res.in}
	
	%% Mark off the abstract in the ``abstract'' environment. 
	\begin{abstract}
	Measurements of the magnetic field strength in the solar corona at heliocentric distances $r\,{\gtrsim}\,1.1\,R_{\odot}$, using observations other than transient radio bursts, are rare. Circular polarization observations of thermal bremsstrahlung radio emission at frequencies $<$\,100\,MHz, which typically originate at $r\,{>}\,1.1R_{\odot}$ in the solar corona, can be effectively used for this purpose. The thermal bremsstrahlung emission is randomly polarized, but in the presence of a magnetic field it propagates in two oppositely polarized circular modes whose differential absorption in the medium results in an observable circular polarization. We report a case study of two-dimensional imaging observations of the above mentioned circular polarization in the solar corona at 51\,MHz using the Gauribidanur RAdioheliograPH (GRAPH), and estimates of the associated coronal magnetic field strength. The peak degree of circular polarization ($dcp$), calculated from the observed total and circularly polarized intensity images with the GRAPH, is ${\approx}$\,2.7\,\%. Using ray-tracing simulations we find that the observed peak $dcp$ corresponds to a radio source located at $r\,{\approx}\,1.65\,{\pm}\,0.15\,R_{\odot}$ at 51\,MHz, and its field strength is ${\approx}\,1.55\,{\pm}\,0.3\,$Gauss.
	\end{abstract}
	%%
	%% You can use the \uat command to link your UAT concepts back its source.
	
	\section{Introduction}\label{sec:intro}
	As the magnetic field plays a key role in the dynamics of the solar corona (see, e.g. \citealp{Aschwanden_2005}), estimates of coronal magnetic field strength ($B$) are very essential to understand the activities in the corona. The related techniques and results for the `inner' corona ($r\,{<}\,1.1\,R_{\odot}$; $R_{\odot}$ is solar photospheric radius)
	%, where $R_\odot$ is Photospheric radius) 
	are widely reported \citep{Kruger1993,Alissandrakis1994,Grebinskij2000,Lin_2004,Zhao2002,Gelfreikh2004,
	Ryabov2004,White2004,Lee2007,Alissandrakis_2021,Tan2022}. Radio observations play a major role in these measurements. Compared to the above, similar estimates of $B$ in the `middle' corona ($r\,{\approx}\,1.1\,-\,3.0\,R_{\odot}$), particularly using observations of thermal radio emission, are rare \citep{Smerd1950,Ramesh2005,Ramesh2010,Sastry_2009,McCauley_2019}. Measurements of $B$ in other regions of the electromagnetic spectrum like ultraviolet and infrared are also very limited \citep{Fineschi1999,Lin_2000,Yang2024}. Extrapolation of the observed magnetic field in the photosphere using Potential Field Solar Surface (PFSS) techniques is a possibility to estimate $B$ in the above-mentioned distance range, but there are assumptions related to the location of the `source surface', currents in the corona, etc \citep{Schatten1969,Schrijver2003}. %In the radio domain, emission from the 
	Radio emission from the Sun at low frequencies when there are no transient bursts, is primarily due to thermal bremsstrahlung and the emission is incoherent and randomly polarized (see, e.g. \citealp{Dulk_1985}). The solar corona is birefringent and anisotropic in the presence of a magnetic field. The randomly polarized thermal bremsstralung radio emission propagates in two oppositely polarized circular modes, the ordinary mode ($O$-mode) and extraordinary mode ($E$-mode). The two modes have different refractive indices, absorption coefficients, and directions of wave propagation, depending on the magnetic field and electron density distributions (see, e.g. \citealp{John_M_Kelso_1964}). So, the total optical depths will be different for the two modes. Hence, there will be a net circular polarization in the observed radio emission from the solar corona \citep{Gary2023}. Determining this net circular polarization w.r.t to the total intensity of the received radio emission could be used to estimate the magnetic field strength responsible for it. Using ray-tracing techniques, \citet{Golap_&_Sastry_1994} and \citet{Sastry_2009} have pointed out the possibility of estimating the magnetic field strength in the corona from the degree of circular polarization ($dcp$) in the observed thermal radio emission at low frequencies. The results of the above authors indicate that the possibility to detect such $dcp$ is comparatively higher at frequencies ${<}$\,100\,MHz. The method is not applicable at frequencies and regions where the corona is optically thick for thermal radiation. The Gauribidanur RAdioheliograPH array (GRAPH; \citealp{Ramesh_1998, Ramesh_2014}) has been regularly observing Stokes-I emission from the solar corona in the above frequency range. Recently, the array was augmented to detect and image circular polarized intensity (Stokes-$V$) also, simultaneously with the total intensity (Stokes-$I$). The present work discusses a case study of Stokes-$I$ and Stokes-$V$ images of thermal emission from the solar corona using the augmented GRAPH array, and estimates of the associated magnetic field strength.
	
	\section{Observations} \label{sec:obs_results}
	\noindent
	\begin{table*}
		\centering
		\caption{Specifications of the augmented GRAPH}
		\label{tab:array_parameters}
		\renewcommand{\arraystretch}{1.5} % Keeps your comfortable vertical spacing
		\begin{tabular}{|c|c|}
			\hline
			\textbf{Parameter} & \textbf{Value}\\
			\hline
			Basic receiving element & Log-periodic dipole array (LPDA) antenna\\
			\hline
			Half-power beam width of LPDA & E-plane: $80^\circ$; H-plane: $120^\circ$\\ 
			\hline
			Operating frequency range of LPDA & 40\,-\,150 MHz\\
			\hline
			Total no. of LPDAs & 512\\
			\hline
			Number of LPDAs per group & 8\\
			\hline
			Array configuration & `T' shaped; East-West (EW) and North-South (NS) arms\\
			\hline
			Number of groups & 64 (32 in EW arm; 32 in NS arm)\\
			\hline
			Orientation of LPDAs in EW arm & $90^o$ in all 32 groups\\
			\hline
			Orientation of LPDAs in NS arm & $90^o$ in 16 groups; $0^o$ in 16 groups\\
			\hline
			Distance between adjacent LPDAs in a group & 10\,m (EW arm); 7\,m (NS arm)\\
			\hline
			Distance between the phase centers of adjacent groups & 80\,m (EW arm); 56\,m (NS arm)\\
			\hline
			Effective collecting area & $192{\lambda}^2$ (Independently for Stokes-$I$ \& Stokes-$V$)\\
			\hline
			Angular resolution & $2.7^{\prime} \times 4.0^{\prime}$ (RA $\times$ Dec.) at 150 MHz\\
			\hline
			Temporal resolution & 256\,msec\\
			\hline
			Observation bandwidth & 1\,MHz\\
			\hline
		\end{tabular}
	\end{table*}
	
	\begin{figure*}
		\centering
		\includegraphics[width=0.9\textwidth]{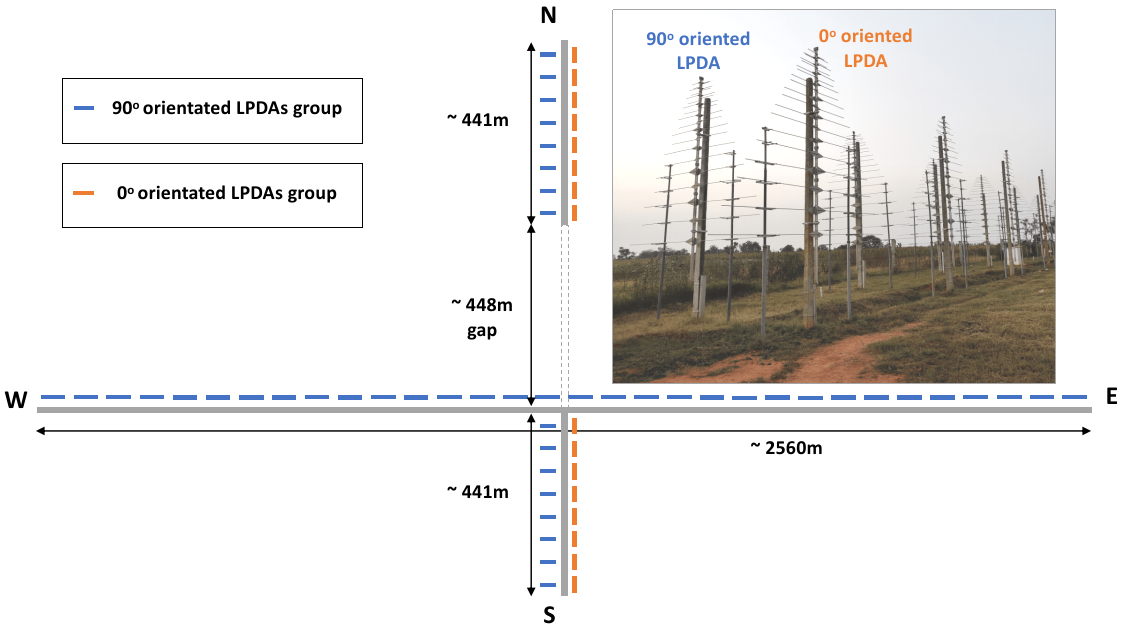}
		\caption{Layout of the augmented GRAPH array (not to scale). Each blue line in the EW arm represents group of 8 LPDAs in 90\textsuperscript{o} orientation (w.r.t the celestial north). In the NS arm, each blue line represents groups of 8 LPDAs in 90\textsuperscript{o} orientation, and each orange line represent groups of 8 LPDAs in 0\textsuperscript{o} orientation (w.r.t the celestial north). The photograph of a small section of South-arm of augmented GRAPH is shown in upper right coroner.}
		\label{fig:array_layout}
		\vspace{0.5cm}
	\end{figure*}
\begin{figure*}
		\centering
		\begin{subfigure}{0.48\textwidth}
			\centering
			\includegraphics[width=\textwidth]{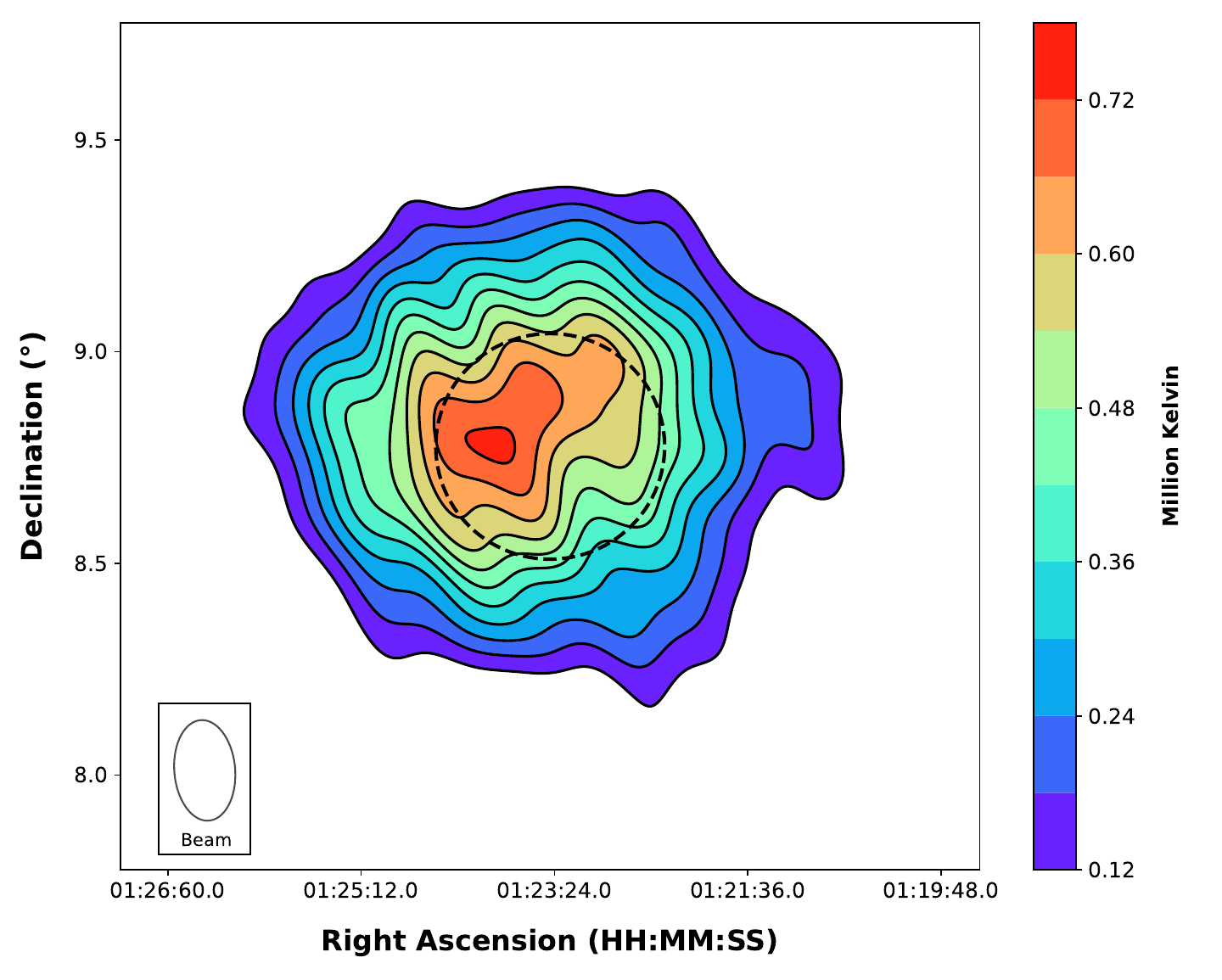}
			\caption{Stokes-I}
			\label{fig:sunicor}
		\end{subfigure}
		\hfill % Adds horizontal space to separate the two images
		\begin{subfigure}{0.48\textwidth}
			\centering
			\includegraphics[width=\textwidth]{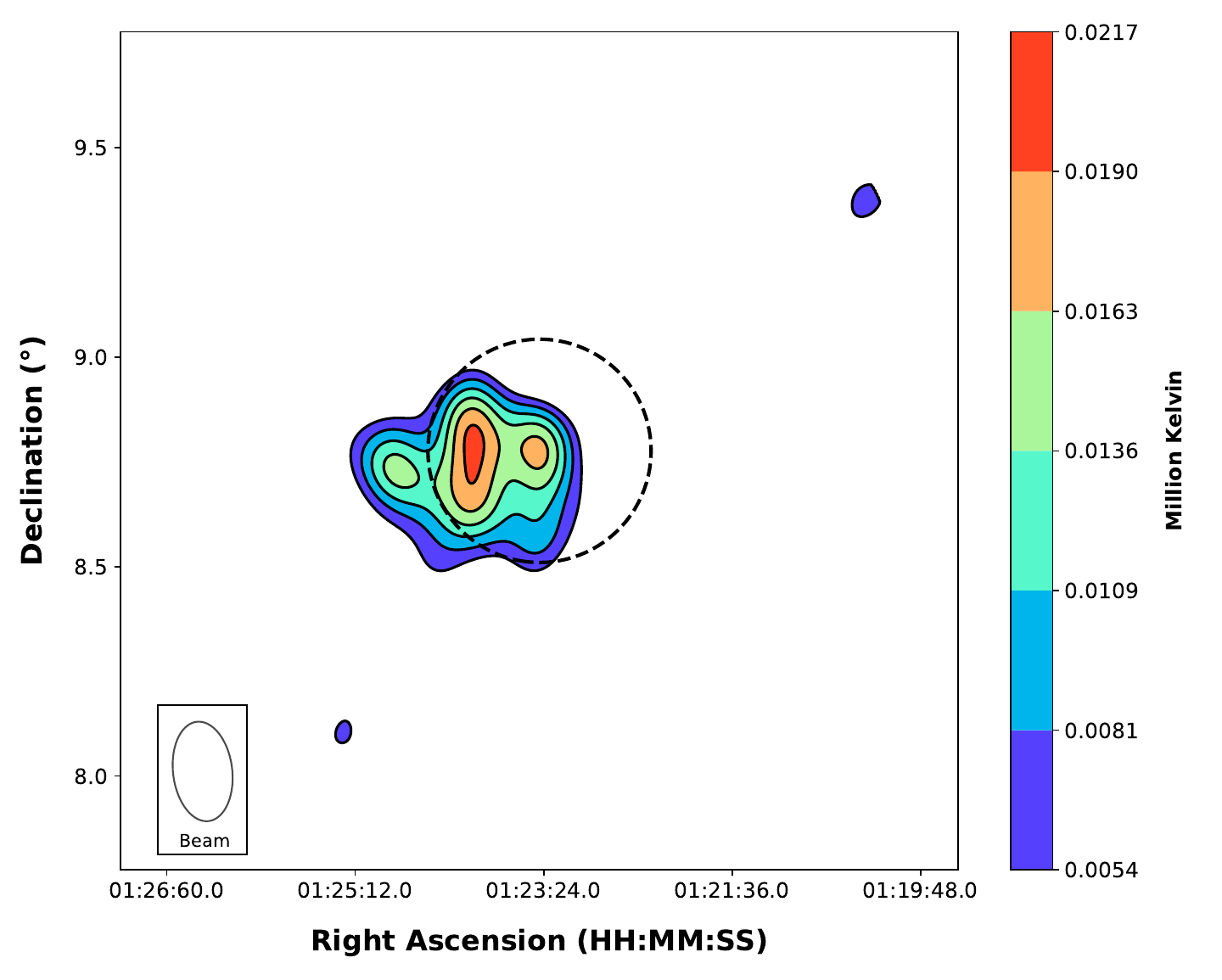}
			\caption{Stokes-V}
			\label{fig:sunvcor}
		\end{subfigure}
		\caption{Solar coronal images observed using the augmented GRAPH on 2025 April 12  at 51\,MHz. The observations were during the local meridian transit of the Sun at ${\approx}$\,06:54\,UT on that day. The black dashed circle in the images represent the solar disk ($1\,R_{\odot}$ radius circle).
		The peak $T_{b}$ in the Stokes-$I$ image is ${\approx}\,7.3\,{\times}\,10^5$\,K. The corresponding value in the Stokes-$V$ image is ${\approx}\,0.2\,{\times}\,10^5$\,K. The estimated $dcp$ from the above two values is ${\approx}$\,2.7\,\%.}
		\label{fig:sunivcor}
		\vspace{0.5cm}
	\end{figure*}

	The GRAPH is a two-dimensional radio interferometer array for dedicated observations of the solar corona in the frequency range 40\,-\,150\,MHz. It is in operation at the Gauribidanur Radio Observatory (GRO; Longitude: 77.44\textsuperscript{o} E \& Latitude: 13.60\textsuperscript{o} N), situated about 100 km north-west of Bangalore (Karnataka, India). In its original configuration, the GRAPH array permitted observations of 
	Stokes-$I$ emission only. It was recently augmented to carry out simultaneous observations of both Stokes-$I$ and Stokes-$V$ radio emission from the solar corona. The specifications of GRAPH are listed in Table \ref{tab:array_parameters} and the array layout is shown in Figure \ref{fig:array_layout}. More details on the augmented GRAPH array and calibration technique can be found in \citet{Sayuf2026b}. The Stokes-$I$ images correspond to the correlations of each group of $90^{\circ}$ oriented LPDAs in the EW arm with each group of $90^{\circ}$ oriented LPDAs in the NS arm. Whereas for the Stokes-$V$, it is the correlations of each group of $90^{\circ}$ oriented LPDAs in the EW arm with each group of $0^{\circ}$ oriented LPDAs in the NS arm.
	Figures \ref{fig:sunicor} and \ref{fig:sunvcor} show the Stokes-$I$ and Stokes-$V$ images obtained on 2025 April 12 at 51\,MHz, during the local meridian transit of Sun at ${\approx}$\,06:54\,UT. 
	No radio bursts were reported\footnote{\url{https://solarmonitor.org/index.php?date=20250412}\label{ftn2}} at the time of this observation. The bandwidth of observation is 1\,MHz, and the data was integrated for 10\,sec to generate the images shown in Figures \ref{fig:sunicor} and \ref{fig:sunvcor}. Observations of Cygnus-A was used for calibration of both the Stokes-$I$ and Stokes-$V$ data, since it is a well known strong calibrator at low radio frequencies \citep{Gasperin2020}, and also unpolarized in our frequency range \citep{Brown1955,Enblin_2019}. The latter implies that presence of any Stokes-$V$ emission in our observations of Cygnus-A is due to instrumental polarization which can be effectively used for calibration (see, e.g. \citealp{Sayuf_grip_2026a}). The data are processed using custom-built software programs and Astronomical Image Processing System (AIPS) software \citep{Wells1985}. The peak observed  brightness temperature, $T_{b}$ of Stokes-$I$ and Stokes-$V$ in Figures \ref{fig:sunicor} and \ref{fig:sunvcor} are 
	${\approx}\,7.3\,{\times}\,10^{5}$\,K and 
	${\approx}\,0.2\,{\times}\,10^{5}$\,K, respectively. The peak $T_{b}$ contours are nearly at the same spatial location in both the Stokes-$I$ \& Stokes-$V$ images, and they closely correspond with the active region AR14060 located at N08E40 on that day$^{\ref{ftn2}}$. The peak $dcp$ of the corresponding discrete radio source is ${\approx}$\,2.7\,\%. 
	 The $T_{b}$ of the observed Stokes-$I$ and Stokes-$V$ emission depends primarily on the electron density ($N\textsubscript{e}$) and magnetic field strength ($B$) near the source region, which can be estimated using ray-tracing simulations \citep{Sastry_2009}.
	
	\section{Ray-tracing simulations} \label{sec:raytrace}
	The use of ray-tracing simulations, in the present context, is to estimate the distribution of $N_{e}$ and $B$ which can reproduce the $T_{b}$ profiles of the observed Stokes-$I$ and Stokes-$V$ emission. Its a kind of back tracing analysis, where $N_{e}$ and $B$ distributions will be created (usually in the form of mathematical expression and as a function of location) and radio wave/ray (of desired frequency) parallel to the Sun-Earth line, initiated from the Earth side towards the Sun, will be traced through the coronal medium till it reaches its closest distance to the Sun in its total ray-path. From the optical depths calculated along the ray-path, the $T_{b}$ of the ray can be calculated. The ray-tracing will be performed for both $O$- \& $E$-modes of rays. The $T_{b}$ corresponding to Stokes-$I$ is calculated by taking an average of the $T_{b}$ of the $O$- \& $E$-modes and, for the $T_b$ of Stokes-$V$, half of the difference between $T_{b}$ of $O$- \& $E$-modes is considered (see Equations \ref{eq:SI} \& \ref{eq:SV}). The $N_{e}$ and $B$ distributions are varied till the $T_{b}$ values for Stokes-$I$ and Stokes-$V$ obtained from the simulations matches the observations. The propagation of a radio ray, for both $O$- \& $E$-modes, in a magnetized plasma medium is described by Haselgrove equations \citep{Haselgrove_1963}. A new ray-tracing python program was developed to simulate the Stokes-$I$ and Stokes-$V$ $T_{b}$, by following the details mentioned in \citet{Haselgrove_1963}, \citet{Golap_&_Sastry_1994}, \citet{kathir_model} and \citet{Sastry_2009}. 
	The mathematical expressions used in the present work for Haselgrove equations, absorption coefficients, refractive indices, etc. are provided in Section \ref{sec:ray_express} and for the $N_e$ and $B$ models in Section \ref{sec:NeB_models}.
	
	\subsection{Formulism}\label{sec:ray_express}
	The direction of the energy related to an electromagnetic radio wave (i.e. poynting vector) travelling in an inhomogeneous anisotropic plasma medium is not parallel to the direction of the wave normal \citep{John_M_Kelso_1964}. This direction of the wave energy is called the ray path. The equations provided by \citet{Haselgrove_1963} help to calculate the rate of change of ray path (see Equation \ref{eq:dxidt}), and the rate of change of wave normal (see Equation \ref{eq:duidt}) for a radio wave in magnetized plasma medium.
	\begin{equation}\label{eq:dxidt}
		\frac{dx_i}{dt}=Ju_i-KYv_i
	\end{equation}
	\begin{equation}\label{eq:duidt}
		\frac{du_i}{dt}=L\frac{\partial X}{\partial x_i}+\sum_j\left(Ku_j+MYv_j\right)\frac{\partial Y_j}{\partial x_i}
	\end{equation}
	\noindent where,
	\begin{eqnarray}
		J&=&2\left[2\left(1-X-Y^2\right)p+Y\left\lbrace 1+\left(v.u\right)^2\right\rbrace \right] \nonumber \\
		K&=&-2X(v.u)(pY-1) \nonumber \\
		L&=&Y\left(1-Y^2\right)p^2-2\left(1-X-Y^2\right)p-Y \nonumber \\
		M&=&2Xp(pY-1) \nonumber  
	\end{eqnarray}
	For a Cartesian coordinate system with the origin at the center of the Sun, first axis towards the Earth, second axis towards the solar west and third axis towards the solar north; the notations are as following:
	\begin{enumerate}
		\item $x_1$, $x_2$, $x_3$ are Cartesian coordinates of a point on the ray path.
		\item $u_1$, $u_2$, $u_3$ are components of $\vec{u}$, parallel to the wave normal at the point ($x_1$,$x_2$,$x_3$). The magnitude of $\vec{u}$ is equal to the refractive index ($\mu$) at that point for the corresponding direction of wave normal (see Equation \ref{eq:ref_indx}).
		\item $v_1$, $v_2$, $v_3$ are components of $\vec{v}$, parallel to the magnetic field at the point ($x_1$,$x_2$,$x_3$).
		\item t is an independent variable.
		%, of no particular significance
		\item $X$\,=\,$(f_p/f)^2$ is a magneto-ionic parameter, where $f_p$ is the plasma frequency and $f$ is the frequency of the radio wave.
		\item $Y$\,=\,$(f_g/f)$ is a magneto-ionic parameter, where $f_g$ is the gyro frequency.
		\item $p$\,=\,$\left(\left|\vec{u}\right|^2-1+X\right)/XY$, obtained from Appleton - Hartree equation (see Equation \ref{eq:ref_indx} and \citealp{Haselgrove_1963}).
	\end{enumerate}
	\noindent Since the plasma medium is considered collision-less (see, e.g. \citealp{Sastry_2009}), the Appleton-Hartree equation for phase refractive index ($\mu$)
	% or $|\vec{u}|$) 
	of a radio wave propagating in a magnetized plasma medium is \citep{Haselgrove_1960,Haselgrove_1963},
	\begin{equation}\label{eq:ref_indx}
		\mu^2=1-\frac{X}{1+(A{\pm}B)}
	\end{equation} 
	\noindent where,
	\begin{itemize}
		\item[] $A$\,=\,$\frac{Y^2\sin^2\theta}{2(1-X)}$
		\item[] $B$\,=\,$\frac{\left[Y^4\sin^4\theta+4Y^2\cos^2\theta\left(1-X\right)^2\right]^\frac{1}{2}}{2\left(1-X\right)}$
		\item[] ${\theta}$ is the angle between the magnetic field and the wave normal, i.e. between $\vec{u}$ and $\vec{v}$.
	\end{itemize}
	\noindent The above mentioned differential equations (\ref{eq:dxidt} \& \ref{eq:duidt}), can be solved using various numerical techniques and in the present work, we have used Runge-Kutta 5th order method.
	
	In our ray-tracing simulation, rays of desired frequency starts from the plane $x_{1}=4\,R_{\odot}$, and are traced towards the Sun. The starting value of $\vec{u}$ is $(-|\vec{u}|,0,0)$. The ray-tracing is done independently for $O$- and $E$-mode rays. Using a step size of $0.001R_{\odot}$, different points on the ray path are calculated till the ray reaches its reflection layer, i.e. the closest reachable layer to the plasma layer corresponding to the frequency of the radio wave used in the calculations). The ray-tracing ends as the ray reaches its reflection layer.  
	At each step in the ray-tracing, optical depth (${\tau}$) is calculated by multiplying the average absorption coefficient in that step region and the corresponding step length of the trace. The total optical depth is obtained by integrating over the entire ray path, and the $T_{b}$ can be calculated for the particular mode. 
	The expressions for the absorption coefficients are,\\
	1) Longitudinal component:
	\begin{equation} \label{eq:klo}
		k_o=\frac{k_n\left(1-X\right)^{1/2}}{\left(1+|Y_l|\right)^{3/2}\left(1-X+|Y_l|\right)^{1/2}}
	\end{equation}
	\begin{equation} \label{eq:kle}
		k_e=\frac{k_n\left(1-X\right)^{1/2}}{\left(1-|Y_l|\right)^{3/2}\left(1-X-|Y_l|\right)^{1/2}}
	\end{equation}
	\\
	2) Transverse component:
	\begin{equation} \label{eq:kto}
		k_o=k_n
	\end{equation}
	\begin{equation} \label{eq:kte}
		k_e=\frac{k_n\left(1+\frac{Y_t^2}{(1-X)^2}\right)}{\left(1-\frac{Y_t^2}{(1-X)}\right)^{3/2}\left(1-\frac{Y_t^2}{(1-X)^2}\right)^{1/2}}
	\end{equation}
	
	\noindent where,
	\begin{itemize}
		\item[] $k_n=\frac{\zeta N_e^2}{f^2\mu T_{ele}^{3/2}}$ is the absorption coefficient in the absence of the magnetic field.
		\item[] $\zeta$ = 0.2 in the corona \citep{Chambe_&_Lantos_1971}.
		\item[] The subscripts `$o$' and `$e$' refer to the $O$- and $E$-modes; `$t$' and `$l$' refer to the transverse and longitudinal components.
	\end{itemize}
	% of the magnetic field.
	The brightness temperature ($T_b$) of a ray in the present case is (see, e.g. \citealp{Dulk1974}),
	\begin{equation}\label{eq:Ta}
		\centering
		T_{b} = T_{ele}\left(1-e^{-\tau}\right)
	\end{equation}
	\noindent The simulated Stokes-$I$ brightness temperature is (see, e.g. \citealp{Alissandrakis_2021}),
	\begin{equation}\label{eq:SI}
		T_{b}^{I}=\frac{T_{b}^{e}+T_{e}^{o}}{2}
	\end{equation}
	
	\noindent where $T_{b}^{o}$ and $T_{b}^{e}$ are brightness temperatures of $O$- and $E$-mode rays. The simulated Stokes-$V$ brightness temperature is,
	\begin{equation}\label{eq:SV}
		T_{b}^{V}=\frac{T_{b}^{e}-T_{b}^{o}}{2}
	\end{equation}
	\noindent The degree of circular polarization is,
	\begin{equation}\label{eq:DCP}
		dcp=\left(\frac{T_{b}^{V}}{T_{b}^{I}}\right)*100
	\end{equation}
	
	\subsection{Coronal electron density and magnetic field models}\label{sec:NeB_models}
	We used the coronal electron density model of \citet{Newkirk_1961}, and the empricial relation for the coronal magnetic field given by \citet{Dulk_&_Mclean_1978} for the present work. These models are considered to hold good in the distance range $r\,{\approx}\,1.1-5\,R_{\odot}$. The rectangular coordinate system is used with Sun's center as origin, `X' direction is towards Earth, `Y' direction is towards the solar west and `Z' direction is towards the solar north. Let x, y \& z be the location co-ordinates (in the units of `$R_\odot$') of a point `P' in the corona. We formulate the electron density ($N_{e}$) at that point as,
	\begin{equation} \label{eq:den}
		N_e(P)=D*4.2*10^4*10^{\frac{4.32}{\rho r}}\left(1+\sum_{i=1}^{n} C_i e^{-\beta_i^2}\right)
		\;cm^{-3}
	\end{equation}
	
	\noindent Similarly, we formulate the magnetic field ($B$) at the same point as, 
	\begin{equation} \label{eq:mag}
		%	\overrightarrow{B(P)}=\frac{B_o}{(r-1)^{\alpha_o}}\left(1+\sum_{i=1}^{n} b_i e^{-%\beta_i^2}\right) \hat{r}\;Gauss
		B(P)=\frac{B_o}{(r-1)^{\alpha_o}}\left(1+\sum_{i=1}^{n} b_i e^{-\beta_i^2}\right)
		\;Gauss
	\end{equation}
	\noindent where,
	\begin{itemize}
		\item[] $r=\sqrt{x^2+y^2+z^2}$ is the heliocentric distance of the point `P' in  units of $R_\odot$
		\item[] ($x_i, y_i, z_i$) is the location of the $i_{th}$ localized source center in units of $R_\odot$
		\item[] $\sigma_{x_i}, \sigma_{y_i}, \sigma_{z_i}$ are the widths of the $i_{th}$ localized source in units of $R_\odot$ along the x, y, z directions, respectively
		\item[] $\beta$ is the width of the localized source;\\
		$\beta^2=\frac{(x-x_i)^2}{2\sigma_{x_i}^2}+\frac{(y-y_i)^2}{2\sigma_{y_i}^2}+\frac{(z-z_i)^2}{2\sigma_{z_i}^2}$
		\item[] $D$ and $B_{o}$ are the density and magnetic field strength enhancement/reduction factors, respectively, for the `background’ corona
		\item[] $C_i$ and $b_i$ are the density and magnetic field strength enhancement/reduction factors, respectively, for the $i_{th}$ localized source
		\item[] $\rho$ and $\alpha_o$ are the factors for the radial variation of $N_{e}$ and $B$ distributions respectively.	
	\end{itemize}
	%\noindent
	%$\hat{r}$ is the unit vector which is radially outwards and parallel to the vector %joining point P and origin (Sun center).
	\noindent The mathematical expressions before the brackets in the Equations \ref{eq:den} \& \ref{eq:mag} can be considered as the density and magnetic field distributions for the background corona, respectively.
	%In the $B$ model we used at these coronal heights, the magnetic field vector is considered to be radially outwards from Sun center \citep{Sastry_2009}}.
	
	\section{Magnetic field strength estimation using Ray-tracing simulations}\label{sec:Bestimate}
	\begin{figure*}
		\centering
		\begin{subfigure}{0.48\textwidth}
			\centering
			\includegraphics[width=\textwidth]{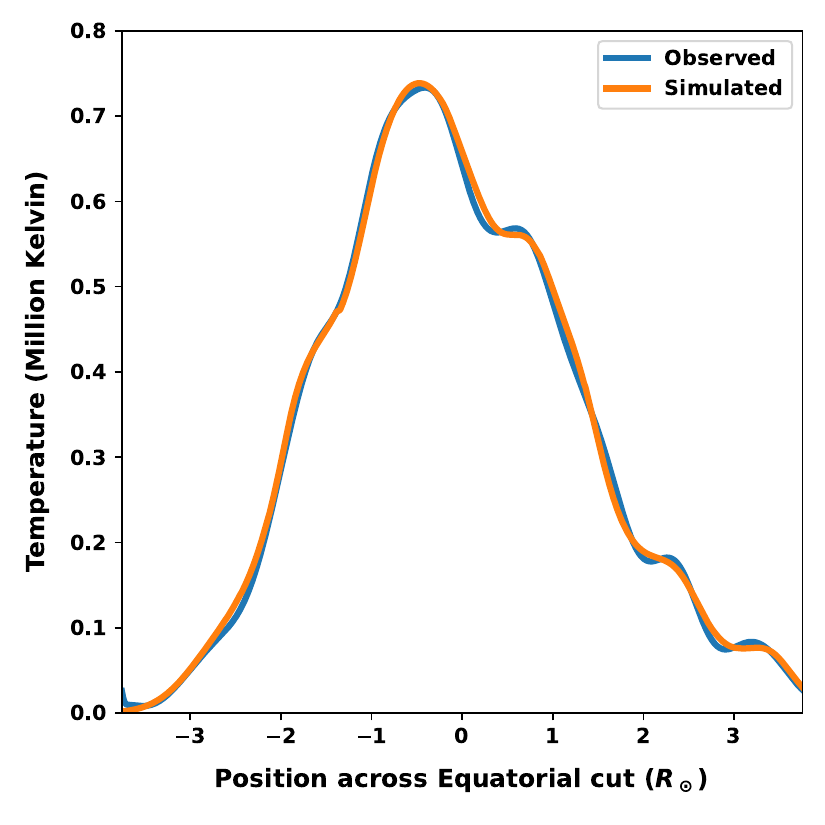}
			\caption{Stokes-I}
			\label{fig:obs_sim_i}
		\end{subfigure}
		\hfill % Adds horizontal space to separate the two images
		\begin{subfigure}{0.48\textwidth}
			\centering
			\includegraphics[width=\textwidth]{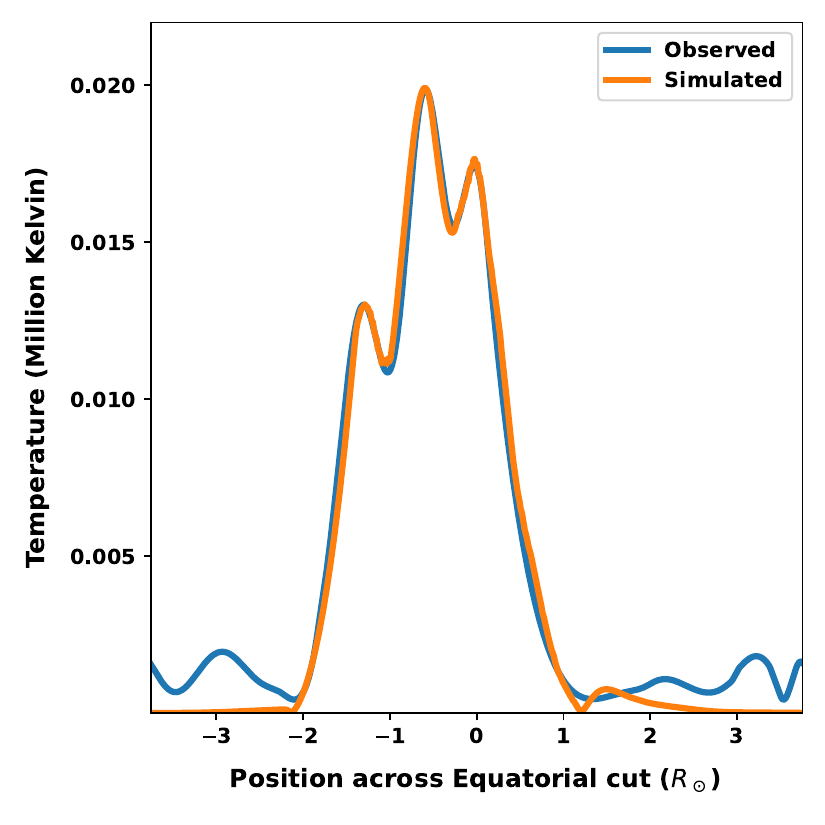}
			\caption{Stokes-V}
			\label{fig:obs_sim_v}
		\end{subfigure}
		\caption{Observed and simulated Stokes-I $T_{b}$ profiles in the equatorial plane of the Sun.}
		\label{fig:obs_sim_iv}
		\vspace{0.5cm}
	\end{figure*}
	
	For the present work, we limit ourselves to ray-tracing simulations of the equatorial brightness distribution. The peak emission in the observed Stokes-$I$ and Stokes-$V$ images in Figures \ref{fig:sunicor} and \ref{fig:sunvcor} are also in the equatorial region. So, we should be able to estimate $N_{e}$ and $B$ around the localised discrete source corresponding to the observed peak emission. Equatorial profiles of the observed two-dimensional Stokes-$I$ and Stokes-$V$ images are considered, and ray-tracing simulations (at 51\,MHz, the observation frequency in the present case) are carried out to simulate the observed equatorial profiles with minimal error. The field-of-view of the images in Figures \ref{fig:sunicor} and \ref{fig:sunvcor} is $2^o \times 2^o$, with a pixel resolution of 14.0625$^{\prime\prime}$. So, the equatorial profile has an extent of ${\approx}\,7.5061\,R_{\odot}$ (i.e. $-3.75305\,R_{\odot}$ to $3.75305\,R_{\odot}$). 
	Figures \ref{fig:obs_sim_i} and \ref{fig:obs_sim_v} shows the simulated and observed Stokes-$I$ and Stokes-$V$ brightness temperature profiles along the equatorial plane, respectively. There is a good correspondence between the observed and simulated profiles. The ray-tracing parameters and the $N_{e}$, $B$ model values used in our simulations are shown in Table \ref{tab:ray_parameter_set}. We used $D$\,=\,0.2 and the 51\,MHz plasma level in the background corona with the above factor will be at ${\approx}\,1.2\,R_{\odot}$, consistent with the observed minimum diameters of the background corona at 53\,MHz and 80\,MHz \citep{Ramesh2020}. 
	Similarly, we used $B_{0}$\,=\,0.1, consistent with $B\,{\approx}$\,600\,mG in the background corona at 1.3\,$R_{\odot}$ mentioned in \citet{Sastry_2009}. Since our simulations are limited to the equatorial plane, the positions of the localised sources along the Z-axis are kept as $0.0R_{\odot}$ ($z_{i}$\,=\,0.0). The source sizes along the Z-axis are given a nominal value of $0.30\,R_{\odot}$ ($\sigma_{z_i}=0.30$), to avoid program execution error. However, these Z-axis parameters will not affect ray-tracing in the equatorial plane.
	\noindent
	\begin{table*}
		\centering
		\caption{Ray-tracing, $N_e$ \& $B$ model parameters used to simulate the Stokes-$I$ and Stokes-$V$ profiles in Figure \ref{fig:obs_sim_iv}}
		\label{tab:ray_parameter_set}
		\renewcommand{\arraystretch}{1} % Provides clean vertical padding
		\setlength{\extrarowheight}{3pt}
		\begin{tabular}{|c|c|}
			\hline
			\centering Frequency, $\mathbf{f}$ & \centering 51\,MHz \tabularnewline
			\hline
			\centering $\mathbf{D}$ & \centering 0.20 \tabularnewline
			\hline
			\centering $\mathbf{B_o}$ & \centering 0.1 \tabularnewline
			\hline
			\centering $\mathbf{\rho}$ & \centering 1.0 \tabularnewline
			\hline
			\centering $\mathbf{\alpha_o}$ & \centering 1.5 \tabularnewline
			\hline
			\centering $\mathbf{T_{ele}}$ & \centering $1{\times}10^{6}$\,K \tabularnewline
			\hline
			\centering $\mathbf{z_{i}}$ & \centering 0.00 (equatorial position) \tabularnewline
			\hline
			\centering $\mathbf{\sigma_{zi}}$ & \centering 0.30 (nominal size) \tabularnewline
			\hline
		\end{tabular}
		
		\vspace{0.5cm} % Adds 1 cm of vertical space between the tables
		\begin{tabular}{|c|c|c|c|c|}
			\hline
			\centering \textbf{Source No.($i$)} & \centering $\mathbf{x_i}$, $\mathbf{y_i}$ & \centering $\mathbf{\sigma_{xi}}$, $\mathbf{\sigma_{yi}}$ & \centering $\mathbf{C_i}$ & \centering $\mathbf{b_i}$ \tabularnewline
			\hline
			\centering 1 & \centering  +1.10, -0.85 & \centering 0.50, 0.50 & \centering +15.00 & \centering +13.00 \tabularnewline
			\hline
			\centering 2 & \centering +1.00, -1.30 & \centering 0.30, 0.20 & \centering +00.00 & \centering +18.00 \tabularnewline
			\hline
			\centering 3 & \centering  +1.30, +0.06 & \centering 0.20, 0.18 & \centering +00.00 & \centering +03.00 \tabularnewline
			\hline
			\centering 4 & \centering +1.00, +1.20 & \centering 0.40, 0.36 & \centering +00.50 & \centering +00.00 \tabularnewline
			\hline
			\centering 5 & \centering +0.80, +2.00 & \centering 0.40, 0.40 & \centering +04.00 & \centering +00.00 \tabularnewline
			\hline
			\centering 6 & \centering +0.60, +2.70 & \centering 0.50, 0.40 & \centering +15.00 & \centering +00.00 \tabularnewline
			\hline
			\centering 7 & \centering +0.60, +3.60 & \centering 0.40, 0.36 & \centering +30.00 & \centering +00.00 \tabularnewline
			\hline
			\centering 8 & \centering +0.80, -1.90 & \centering 0.50, 0.48 & \centering +10.00 & \centering +00.00 \tabularnewline
			\hline
			\centering 9 & \centering +0.60, -3.00 & \centering 0.50, 0.50 & \centering +15.00 & \centering +00.00 \tabularnewline
			\hline
		\end{tabular}
	\end{table*}
 
\begin{figure*}
		\centering
		\begin{subfigure}{0.48\textwidth}
			\centering
			\includegraphics[width=\textwidth]{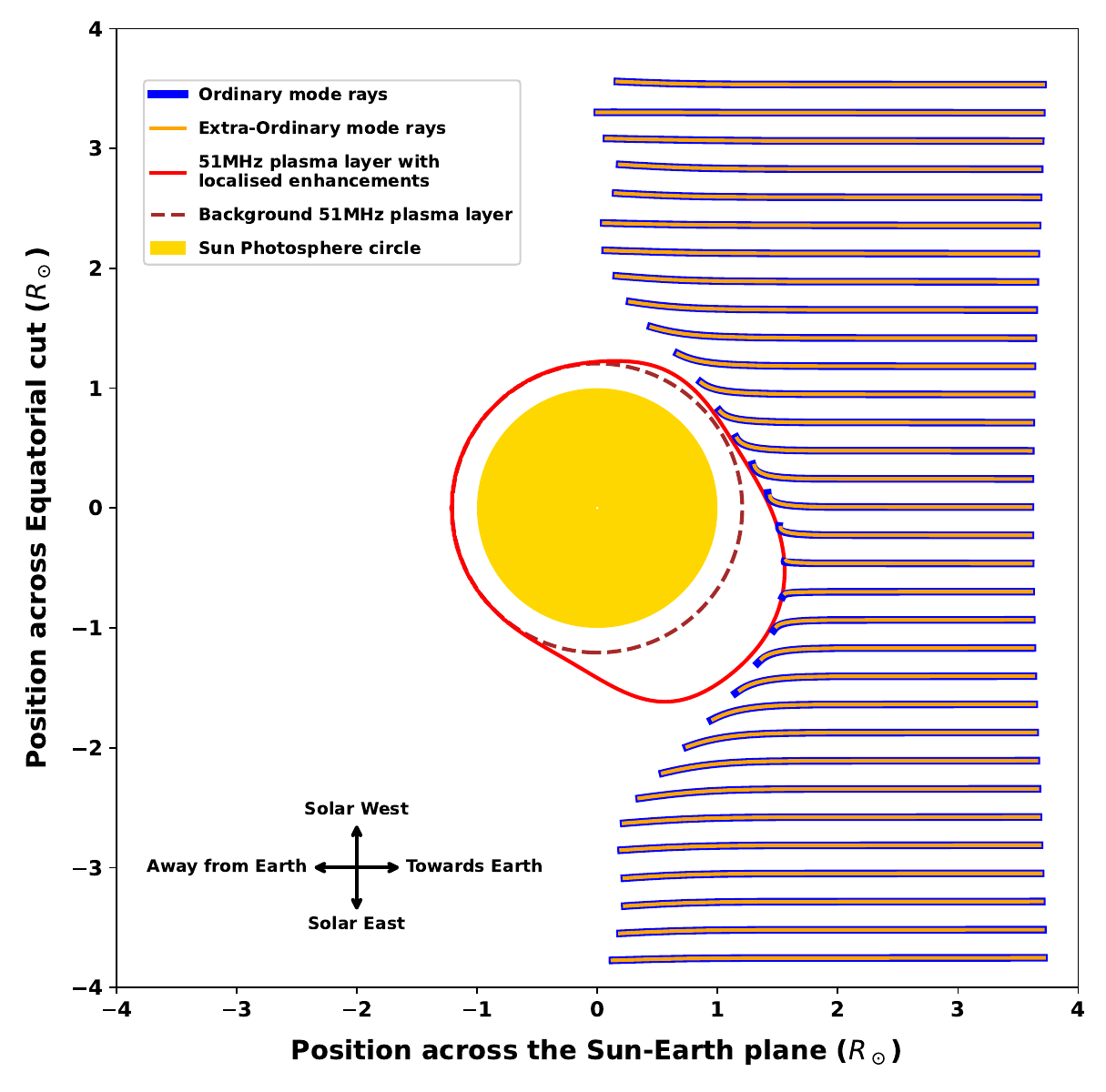}
			\caption{}
			%	\caption{Ray-tracing representation in the Sun-Earth plane. Solar north is %perpendicular to the plane of the paper in the viewing direction. Solar east and %west sides are in the downward and upward directions in the plane of the paper, %respectively. Right side is the Earthward direction.}
			\label{fig:raytraces1}
		\end{subfigure}
		\hfill % Adds horizontal space to separate the two images
		\begin{subfigure}{0.48\textwidth}
			\centering
			\includegraphics[width=\textwidth]{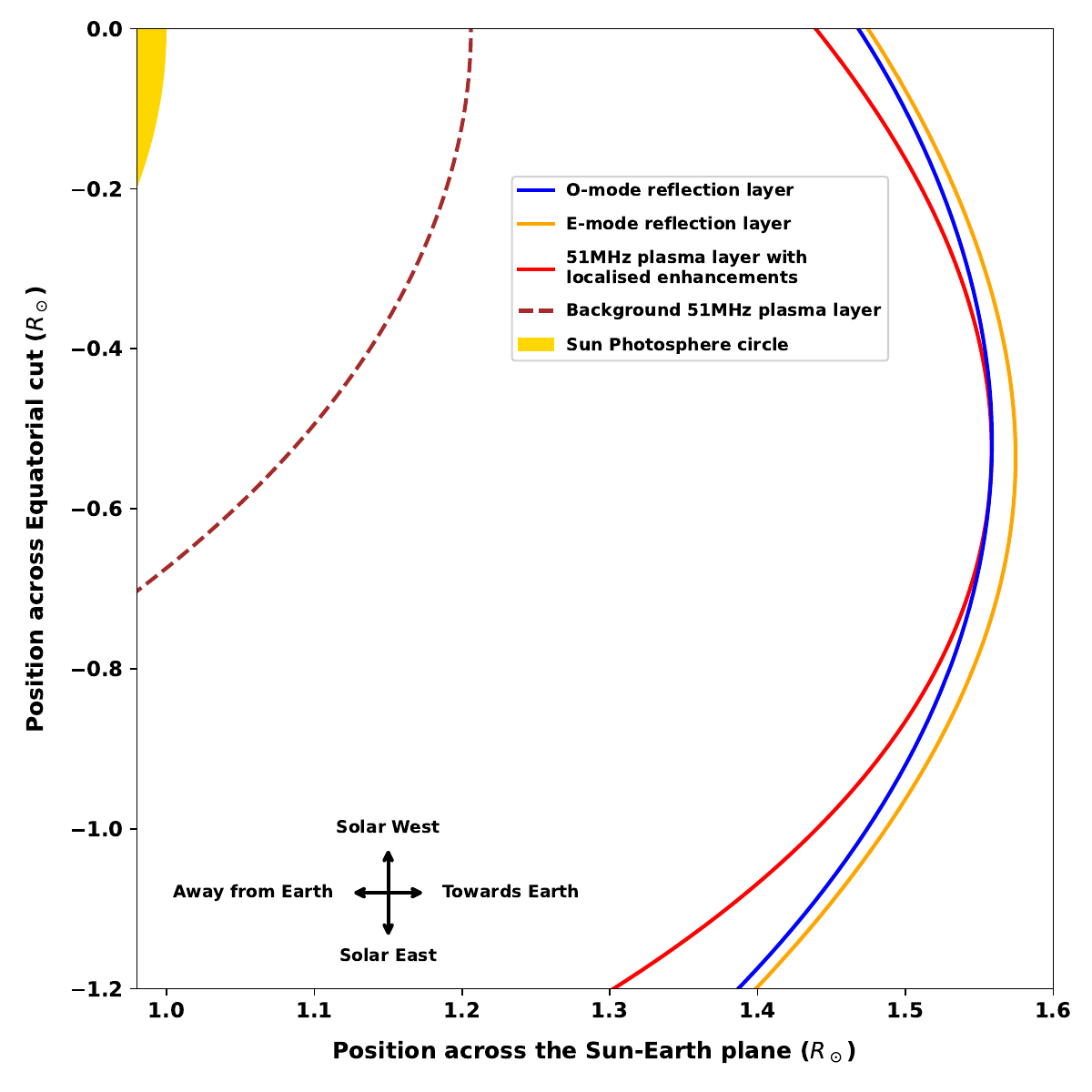}
			\caption{}
			%	\caption{Reflection levels of O- \& E-modes near the location of maximum optical %depth in the present case.}
			\label{fig:reflayers1}
		\end{subfigure}
		\caption{(a) Ray-tracing representation in the Sun-Earth plane. Solar north is perpendicular to the plane of the paper in the viewing direction. 
			%Solar east and west directions are vertically downward and upward in the plane %of the paper, respectively. Right side is the Earthward direction. 
			(b) Reflection levels of O- \& E-modes near the location of maximum optical depth in the present case.}
					\label{fig:raytraces}
	\end{figure*}

	An illustration of the ray-tracing simulation of the present study is shown in Figure \ref{fig:raytraces1}. The view is from the top of the Sun-Earth plane, with Earth on the right hand side.
	%Solar north is to the left, and solar east is vertically downwards. 
	Traces of few $O$- and $E$-mode rays from Earth-ward side towards the Sun-ward side, till their reflection levels are shown. 
	The brown colour `dashed' lines forming a circle of radius 1.2\,$R_\odot$, indicate the location of the 51\,MHz plasma level considering the background corona distributions only. The red colour closed line with a prominent bulging on the lower right is the 51\,MHz plasma level in the presence of localized electron density and magnetic field  enhancements above the background corona (Table \ref{tab:ray_parameter_set}). The localized sources were established via an iterative multi-Gaussian least-squares curve fitting technique taking into consideration the strength and location of the sources w.r.t the background corona (see, e.g. \citealp{Ramesh2006}). The location of the bulging in Figure \ref{fig:raytraces1} correlates with the steady coronal streamers observed in the solar east on 2025 April 12, the same day as the observations in Figure {\ref{fig:sunivcor}} \footnote{\url{https://soho.nascom.nasa.gov/data/REPROCESSING/Completed/2025/c2/20250412/}} and is probably an indication of the extension of a streamer from the associated active region (AR14060 located at N08E40 on that day (Section \ref{sec:obs_results})) into the three-dimensional space (see, e.g. \citealp{Liewer2001}). 
	The angle between the axis of the bulge and the Sun-Earth line in Figure \ref{fig:raytraces1} is ${\approx}\,45^{\circ}$. This closely matches the location of the associated active region mentioned above.
	In the presence of magnetic field, the reflection level of the $O$-mode will be similar to that of the non-magnetic field case, but the reflection level of the $E$-mode will be at a slightly higher height, i.e. above the $O$-mode reflection level. The absorption coefficients will be also different for the two modes (see, e.g. \citealp{Fomichev1968,Alissandrakis_2021}). For the present ray-tracing analysis, the reflection levels of $O$- and $E$-modes, and the 51\,MHz plasma layer, near the location of the peak $T_{b}$, are shown in Figure \ref{fig:reflayers1}.
	%Figure \ref{fig:reflayers}. 
	The reflection level of the $O$-mode
	coincides with the 51\,MHz plasma layer, and the $E$-mode layer is slightly above at a distance of  
	%at that location 
	${\approx}\,0.017\,R_{\odot}$. The separation between the locations of the reflections levels of the $O$- and $E$-modes is a function of the  magnetic field strength (see, e.g. \citealp{Fomichev1968,Tan2022}).  
	Difference between the absorption coefficients of the $O$-mode and $E$-mode, leading to a net $dcp$, will be more for larger longitudinal magnetic field component, i.e. line-of-sight component (see Equations \ref{eq:klo} to \ref{eq:kte}). In other words, sources with larger line-of-sight magnetic field strength will contribute more to the Stokes-$V$ emission (see, e.g. \citealp{Sastry_2009}).
	
\begin{figure*}
		\centering
		\includegraphics[width=0.95\textwidth]{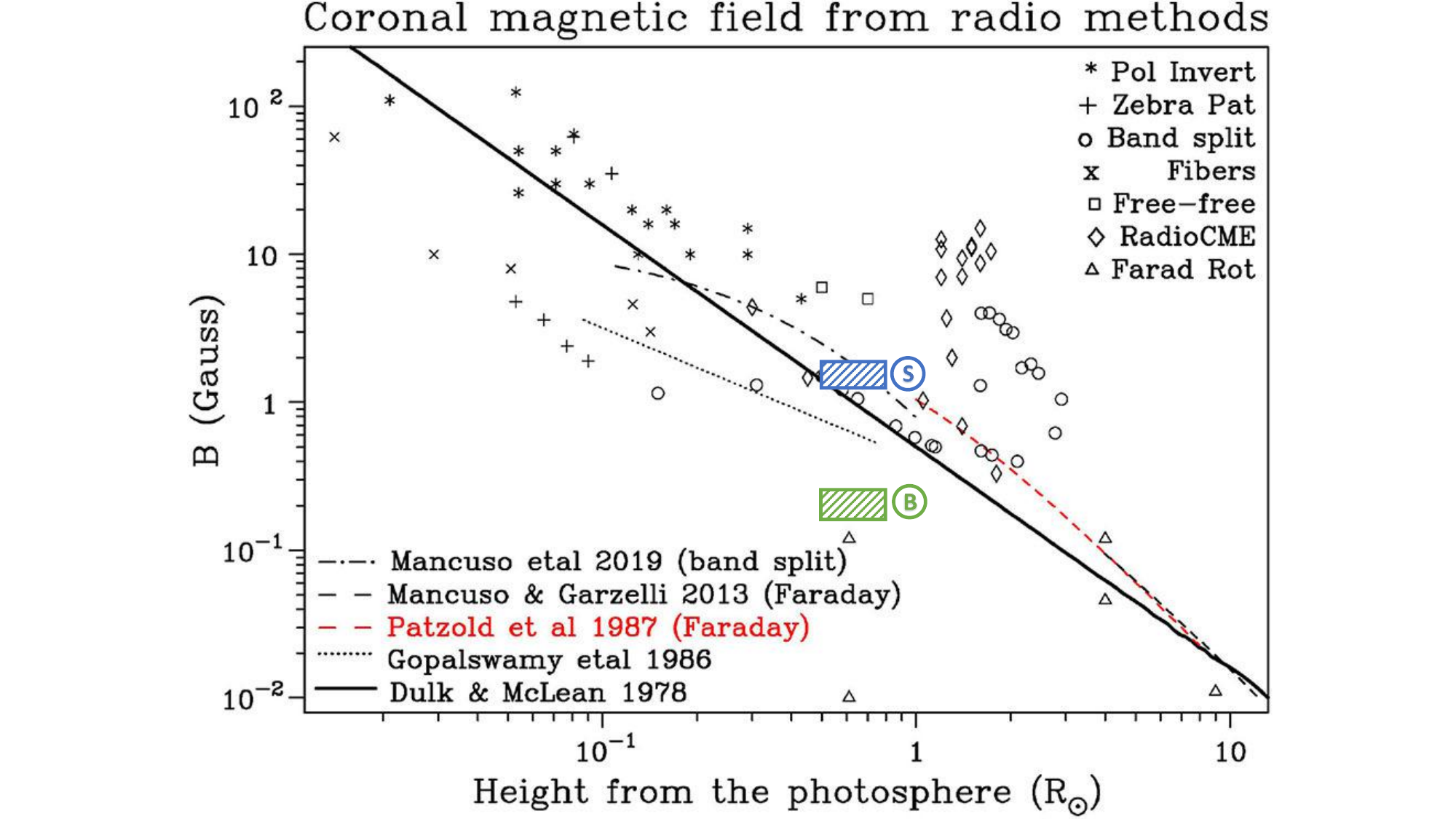}
		\caption{Solar coronal magnetic field strength at different heights in the corona estimated using various radio techniques \citep{Alissandrakis_2021}. The rectangular boxes `S' and `B' represents the magnetic field strengths estimated from the ray-tracing simulations at ${\approx}\,1.65\,R_{\odot}$ in the present work. They correspond to the discrete source (Section \ref{sec:Bestimate}) and the background corona (equation \ref{eq:mag} and Table \ref{tab:ray_parameter_set}), at the above location, respectively. The extent of the boxes indicates the error in the estimates.}
		\label{fig:B_est_comp}
	\end{figure*}

	The magnetic field at the peak $T_{b}$ location 
	$(x,y,z = 1.557,\,{-}0.56,\,0.00)R_{\odot}$ is ${\approx}$\,1.55\,G. The electron density is ${\approx}\,3.2\times10^7\,cm^{-3}$, which corresponds to 51\,MHz plasma frequency. The heliocentric distance corresponding to the $x,y,z$ coordinates is 1.65\,$R_{\odot}$.
	To check the dependency of the results on the $N_{e}$ and $B$ models, we carried out ray-tracing simulations with different values for the parameters $D$, $B_o$, $\rho$, $\alpha_o$, $C_i$, $b_i$, etc. in Table \ref{tab:ray_parameter_set}.
	The variations in the above mentioned heliocentric distance and the $B$ are within 
	${\pm}\,0.15\,R_{\odot}$ and ${\pm}$\,0.3\,G, respectively. Note that the location of the 51\,MHz plasma layer in the background corona of the present $N_e$ distribution differs by a maximum of 0.15\,$R_{\odot}$ w.r.t those in the other coronal electron density models \citep{Baumbach1937,Allen1947,Saito1977}.
	The day-to-day changes in the observed size of the Sun, free of any burst activity, at 51\,MHz and adjacent  frequencies, reported by different authors based on observations at different epochs, are nearly in the same range as above \citep{Gergely1985,Thejappa1994,Ramesh2006,Ramesh2020,Zhang2022}. Furthermore, we find that the above density of ${\approx}\,3.2{\times}10^{7}\,cm^{-3}$ at ${\approx}\,1.65\,R_{\odot}$ in the present case is slightly higher compared to the recent estimates of ${\approx}\,0.76{\times}10^{7}\,cm^{-3}$ for the
	same distance by \cite{Mondal2026} and the mean value of ${\approx}\,0.6{\times}10^{7}\,cm^{-3}$ at the same	distance in the compilation of the density estimates reported by different authors from different types of observations in \cite{Wexler_2020}. The difference could be due to the fact that our estimates corresponds specifically to	the location of the density enhancement at the above distance. For e.g., the density measurements in the vicinity of a streamer at the same	distance mentioned in \cite{Wexler_2019} is ${\approx}\,1{\times}10^{7}\,cm^{-3}$, which is closer to our results and the above mentioned density of $3.2{\times}10^{7}\,cm^{-3}$ falls at ${\approx}\,1.57\,R_{\odot}$ which is within the 0.15\,$R_{\odot}$ range w.r.t the present estimate. The estimated $B$ in our present study agrees reasonably well with previous measurements using other techniques (see, e.g. the review by \citealp{Alissandrakis_2021}). Figure \ref{fig:B_est_comp} shows the comparison. According to \citet{Drago1994}, for a magnetic field of $B\,{\approx}$\,1\,-\,2\,G in a coronal streamer, the expected $dcp$ due to thermal emission is ${\leq}$\,3\,\%. \citet{Ramesh2021} reported $B\,{\approx}$\,0.5\,G at $r\,{\approx}\,2.1\,R_{\odot}$  based on observations of thermal emission from a coronal mass ejection, which too is density enhancement in the corona like a streamer. The present values, i.e. $dcp\,{\approx}$\,2.7\,\% (Section \ref{sec:obs_results}), and $B\,{\approx}$\,1.55\,G, for the discrete radio source associated spatially with an active region, agree closely with the above numbers. We find that a 10${\times}$ density increase to the 0.2${\times}$\citet{Newkirk_1961} model assumed for background corona, in the present case, indicates that 51\,MHz plasma level would be at ${\approx}1.67\,R_{\odot}$, which is fairly consistent with the heliocentric distance corresponding to the location of the peak $T_{b}$ mentioned earlier. Since dense streamers are considered to be 10${\times}$ more denser than the background corona (see, e.g. \citealp{Thejappa_&_kundu_1992}), it is likely that the active region associated discrete source mentioned above corresponds to a streamer source (see, e.g. \citealp{Dulk1974}). \\ 

	The magnetic field distribution model used in the present work to match the observed Stokes-$I$ \& Stokes-$V$ profiles indicates that the field strength in the background corona is ${\approx}\,0.2\,{\pm}\,0.04$\,G at 1.65\,$R_\odot$ (green box in Figure \ref{fig:B_est_comp}). Interestingly, \cite{Wexler_2019} had earlier estimated $B\,{\approx}$\,0.12\,G at 1.7\,$R_\odot$ in a non-active region of the equatorial corona, using spacecraft Faraday rotation technique. There is a close agreement between the two independent measurements, carried out using different techniques and at different epochs. This suggests that the strength of the magnetic field in the background corona can be low as 0.1\,-\,0.2\,G around ${\approx}\,1.7R_{\odot}$. Employing similar technique, \cite{Wexler_2021} estimated $B\,{\approx}$\,1.18\,G at 1.64\,$R_\odot$, for a coronal structure with closed magnetic field. Our present results for thermal emission from the discrete coronal source in our observations (Figure \ref{fig:obs_sim_iv}) is in reasonable agreement in this case also. The $B$ values at the location of the discrete sources is ${\approx}$\,7\,-\,10${\times}$ larger compared to the background corona. This suggests that the assumption of `mean $B$' may not be strictly applicable at distances like $r\,{\approx}\,1.65\,R_{\odot}$ as in the present work (see, e.g. \citealp{Wexler_2019}). Equation \ref{eq:mag} and Table \ref{tab:ray_parameter_set} indicates the same. Note that we used $T_{ele}\,{=}\,10^{6}$\,K for our present calculations (Table \ref{tab:ray_parameter_set}), the typical value for the background in the `middle' corona. However, $T_{ele}$ can be $1.4{\times}10^{6}$\,K and $1.8{\times}10^{6}$\,K in the regions of fast and slow solar winds, respectively, at 1.5\,$R_{\odot}$ (see, e.g. \citealp{West2023}). 
	According to \citet{Boe2020}, $T_{ele}$ is limited to $1.4{\times}10^{6}$\,K even in streamers, potential source region for the slow solar wind. In such case of $T_{ele}\,{\approx}\,1.4{\times}10^{6}$\,K , our calculations indicate $B\,{\approx}\,1.13\,{\pm}\,0.3$\,G at ${\approx}\,1.65\,R_{\odot}$, which is in excellent agreement with the results of \citet{Wexler_2021} mentioned above. 
	%-----------
	
	\section{Summary} \label{sec:summary}
	We have presented Stokes-$I$ and Stokes-$V$ images of thermal emission from the solar corona solar corona at 51\,MHz on 2025 April 12 using observations with the augmented Gauribidanur radioheliograph. The close spatial correspondence between the peak Stokes-$I$ and Stokes-$V$ emission indicate that common localised sources are responsible for the observed emission in both the cases. 
	The peak $dcp$ is 2.7\,\%. To estimate the coronal magnetic field strength 
	%and electron density 
	associated with the observed peak $T_{b}$ in the  
	%Stokes-$I$ and 
	Stokes-$V$ image, one-dimensional ray-tracing calculations in the equatorial plane of the Sun were performed using a python language based ray-tracing software code. The estimated field strength is ${\approx}\,1.55\,{\pm}\,0.3$\,G at ${\approx}\,1.65\,{\pm}\,0.15\,R_{\odot}$. Note that \citet{Ramesh2010} had earlier reported $B\,{\approx}$\,6\,-\,5\,G in the distance range ${\approx}$\,1.5\,-\,1.7\,$R_{\odot}$ from similar observations of thermal emission from streamer sources at 109\,MHz and 77\,MHz. Since the observing frequencies are higher than in the present case, it is likely that their observations correspond to higher density and/or magnetic field in the source region. Two-dimensional simulations of the Stokes-$I$ and Stokes-$V$ brightness temperature distributions is in preparation and shall be reported separately. The results indicate that high dynamic range Stokes-$I$ and Stokes-$V$ observations of thermal emission from the solar corona at low frequencies ($<$\,100\,MHz) with large antenna arrays like the LOFAR, OVRO-LWA, upcoming SKA-Low, etc., and advanced ray-tracing programs using Artificial Intelligence \& Machine Learning (AI-ML) to simulate the observed images could help to estimate the magnetic field strengths in the `middle' corona on a daily basis. 
	
	%% Please use the acknowledgment and contribution environments. This will 
	%% be anonomyized when the "anonymous" style option is used. 
	\begin{acknowledgments}
		We are grateful to the Gauribidanur Observatory team for their help and contribution in the observations and upkeep of the facilities. We also sincerely thank the reviewer whose valuable comments and insights helped us in improving the manuscript. 
	\end{acknowledgments}
	
	%\begin{contribution}
	%%%This section gives authors the space to recognize author contributions. The text inside this environment is NOT counted towards the total word quanta. At a minimum, manuscripts are expected to include this text:
	%
	%\textcolor{red}{contribution of Authors}
	%
	%\end{contribution}
	
	%% To help institutions obtain information on the effectiveness of their 
	%% telescopes the AAS Journals has created a group of keywords for telescope 
	%% facilities.
	\facilities{Gauribidanur Radio Observatory (GRO), Gauribidanur Radioheliograph (GRAPH)}
	
	%% Similar to \facility{}, there is the optional \software command to allow 
	%% authors a place to specify which programs were used during the creation of 
	%% the manuscript. Authors should list each code and include either a
	%% citation or url to the code inside ()s when available.
	\software{Astronomical Image Processing System (AIPS), Python programming language along with Python packages - Numpy, Numba, Sympy, Matplotlib, Astropy, Sunpy and Math.}
	
	%% For this sample we use BibTeX plus aasjournalv7.bst to generate the
	%% the bibliography. The sample7.bib file was populated from ADS. To
	%% get the citations to show in the compiled file do the following:
	%%
	%% pdflatex paper1.tex
	%% bibtext paper1
	%% pdflatex paper1.tex
	%% pdflatex paper1.tex
	
	\FloatBarrier
	
	\bibliography{paper1}{}
	\bibliographystyle{aasjournalv7}
	
	%% This command is needed to show the entire author+affiliation list when
	%% the collaboration and author truncation commands are used.  It has to
	%% go at the end of the manuscript.
	%\allauthors
	
	%% Include this line if you are using the \added, \replaced, \deleted
	%% commands to see a summary list of all changes at the end of the article.
	%\listofchanges

\end{document}